\documentclass[sigconf, nonacm]{acmart}
\usepackage{amsmath,amsfonts}
\usepackage{algorithmic}
\usepackage{balance}
\usepackage{csquotes}
\usepackage{graphicx}
\usepackage{textcomp}
\usepackage{xcolor}
\usepackage{url}
\usepackage{threeparttable}
\usepackage{booktabs}
\usepackage{dcolumn}
\usepackage{makecell}
\usepackage{multirow}
\usepackage{enumerate}
\usepackage[inline]{enumitem}

\AtBeginDocument{%
  \providecommand\BibTeX{{%
    \normalfont B\kern-0.5em{\scshape i\kern-0.25em b}\kern-0.8em\TeX}}}

\setcopyright{acmcopyright}
\copyrightyear{2023}
\acmYear{2023}
\acmDOI{XXXXXXX.XXXXXXX}

\acmConference[]{}{}{}
\acmBooktitle{} 
\acmPrice{}
\acmISBN{}

\begin{document}

\title{Towards An Empirical Theory of Ideologies in the \\Open Source Software Movement}


\author{Yang Yue}
\affiliation{%
  \institution{University of California, Irvine}
  % \streetaddress{1 Th{\o}rv{\"a}ld Circle}
  % \city{Hekla}
  \country{USA}}
\email{y.yue@uci.edu}

\author{Yi Wang}
\affiliation{%
  \institution{Beijing University of Posts and Telecomunicaions}
  % \city{Rocquencourt}
  \country{China}}
  \email{wang@cocolabs.org}

\author{David Redmiles}
\affiliation{%
 \institution{University of California, Irvine}
 \country{USA}}
 \email{redmiles@ics.uci.edu}


\begin{abstract}
  Encompassing a diverse population of developers, non-technical users, organizations, and many other stakeholders, open source software (OSS) development has expanded to broader social movements from the initial product development aims. Ideology, as a coherent system of ideas, offers value commitments and normative implications for any social movement, so does OSS ideology for the open source movement. However, the literature on open source ideology is often fragile, or lacking in empirical evidence. In this paper, we sought to develop a comprehensive empirical theory of ideologies in open source software movement. Following a grounded theory procedure, we collected and analyzed data from 22 semi-structured interviews and 41 video recordings of Open Source Initiative (OSI) board members' public speeches. An empirical theory of OSS ideology emerged in our analysis, with six key categories: membership, norms/values, goals, activities, resources, and positions/group relations; each consists of a number of themes and subthemes. We discussed a subset of carefully selected themes and subthemes in detail based on their theoretical significance. With this ideological lens, we examined the implications and insights into open source development, and shed light on the research into open source as a social-cultural construction in the future.
\end{abstract}





\settopmatter{printacmref=false}
\maketitle

\section{Introduction}

Open source software (OSS) is not only just a software development paradigm \cite{FellerF00FrameworkParadigm} but also a social movement \cite{Crowston12} since OSS realizes the sociological ``coming together'' of various types of individuals and organizations to form a collective identity for certain purposes on Internet-based platforms \cite{diani1992concept, diani2000social, van2010internet}. It brings not only substantial changes in the software industry \cite{Fitzgerald06}, but also has profound implications beyond the technical realm \cite{diani1992concept}. Its scalability, continuous improvements, and community-driven innovations, offer great opportunities to champion social good and help address many challenging issues such as climate change \cite{malone2007harnessing}, human trafficking \cite{kejriwal2017flagit}, pandemic outbreaks \cite{zastrow2020open}, etc. 

Understanding a social movement should not neglect its underpinning ``heart and soul'' \cite{thompson2016society}. Such ``heart and soul'' is exactly the \textbf{ideology} guiding its membership and members' actions, its issue and agenda selections, its ideas about solutions to problems, and its choice of tactics \cite{king2008ideology}. Ideology has been recognized for its critical role in driving such social movements \cite{mccright2008the, valocchi1996the, zald2000ideologically}. For example, according to Leveille's case study \cite{leveille2017searching}, the dynamic of the Occupy Wall Street (OWS) movement between the fall of 2011 and 2012 was determined by the movement's ideological orientations, which also connected an inner core and an outer range of more or less active participants. Similar to any other social movements, the open source movement in the last several decades is also driven by its own ideology, namely, OSS ideology.

OSS ideology could be defined as ``\textbf{the basis of social representations regarding open source development shared by open source community}'' \cite{YueYYWR21}. These social representations enable OSS Ideology to influence the OSS movement at multiple micro-, meso- and macro-levels. At the micro-level, ideology influences individuals' beliefs, preferences, decisions, and actions \cite{chen2006relation}, such as choosing particular open source projects to contribute to, and when to leave a project. Then, at the project and community (meso-) level, OSS ideology also influences a team's governance, participation, and dynamics \cite{StewartG06EffectivenessIdeology, DanielMCH18Misfit}, e.g., why a project has a BDFL--Benevolent Dictator For Life, and how to achieve high productivity with voluntary contributions. Finally, at the broader societal (marco-) level, OSS ideology accounts for the emergence of open innovation systems \cite{beck2022open,huizingh2011open}, or the harnessing of collective intelligence in addressing mission-critical global issues \cite{deek2007open}. Thus, developing a deep understanding of OSS ideology would inform us how the OSS movement has grown and exhibited diversities in its practices while maintaining its core premises; and provide an evaluative framework for identifying and assessing the potential misfits among multiple participating entities' ideological orientations \cite{bermiss2018ideological}.  

While the impacts of OSS ideology are ubiquitous in almost every aspect of the OSS movement, a striking fact is a relatively little research focusing on it in SE literature, in contrast to increasingly investigated many other issues in OSS \cite{YueYYWR21}. Ideological elements often dispense in the literature studying motivation \cite{Gerosa21motivation, sharp2009models}, collaboration and coordination \cite{MockusFH02Coordination, JensenS05}, in a fragile manner. One reason for the dearth of such studies might be that the term ``ideology'' is inherently vague in its conceptualization. Researchers often took a convenient way of focusing on a very narrow proportion of it, e.g., the ideology of effectiveness \cite{StewartG06EffectivenessIdeology}. Moreover, applying an ideological lens to investigate the OSS movement also suffered from the absence of agreed-upon essential elements and an empirically-grounded theoretical framework overarching these elements \cite{YueYYWR21}. 

Aiming to fill the above gaps, this paper reported on our efforts in developing a comprehensive, contemporary, and empirically-grounded theory of OSS ideology. The theory development process started by reviewing the literature in ideology (\textbf{see \S2}), which would be used to guide the following inquiries. We designed a qualitative grounded theory \cite{charmaz2006constructing, muller2010grounded} study to build the theory (\textbf{see \S3}). During the entire grounded theory process, we interviewed 22 OSS participants with diverse backgrounds and compiled a dataset containing 41 Open Source Initiative \footnote{OSI was established in 1998 as a steward organization for the OSS movement.} (OSI)'s current and former members' most recent public speeches/interviews. The data were analyzed and interpreted in an iterative coding and memo-writing procedure. Finally, a substantive theory, as a set of themes that are related to one another under the cohesive umbrella of the OSS ideology, emerged from the data (\textbf{see \S4}). It consisted of 42 themes in six broader categories: \emph{membership}, \emph{norms/values}, \emph{goals}, \emph{activities}, \emph{position and group-relations}, and \emph{resources}. We further discussed the theoretical and practical implications of the theory (\textbf{see \S5}). Our study thus contributed to the SE literature in three ways:

\begin{enumerate}[leftmargin=.68cm, labelsep=-0.2cm, align=left]
\item \emph{A comprehensive, contemporary, and empirically-grounded theory of OSS ideology under a cohesive framework that reflects the current OSS movement};
\item \emph{A collection of detailed descriptions of selective ideological themes offering critical empirical and theoretical insights and/or opportunities for future research};
\item \emph{A set of implications regarding insights into the OSS movement, future research opportunities, and practices, with the ideological lens offered by our empirical theory}.
\end{enumerate}

%The rest of the paper proceeds as follows: Section \ref{related_work} introduces a brief history of the concept of ideology, and a few studies related to OSS ideology. Section \ref{definition} presents the definition of OSS ideology, and Section \ref{methods} introduces our research methods. The empirical theory of OSS ideology is presented in Section \ref{theory}, while tensions identified in the analysis are presented in Section \ref{tension}. Some implications are discussed in Section \ref{discussion}, and Section \ref{conclusion} concludes the paper.

\section{Background and Related Work}
\label{related_work}

\subsection{A Brief History of Ideology}

%\begin{figure}[h]
%    \centering
%    \includegraphics[width=\columnwidth]{Timeline of Ideology Theory.pdf}
%    \caption{The simplified key fundamental developments of ideology theories}
%    \label{fig:timeline}
%\end{figure}

The concept of ideology is historically vague and ambiguous \cite{VanDijk1998}. 
%Fig. \ref{fig:timeline} is the timeline of some major progresses of ideology theories. 
The term ``ideology'' was first used by French philosopher Destutt de Tracy in 1796, when he started a project to analyze ideas and sensations. He defined ideology as the ``science of ideas,'' which was a positive and rigorous concept in science \cite{KENNEDY1982}. Then he extended the concept's scope to social and political disciplines and focused on ideas themselves, i.e., experience, feeling, etc. Therefore, ideology gradually became abstract and illusory ideas \cite{thompson2013ideology}.

Karl Marx played a crucial role in developing ideology theory \cite{VanDijk1998}. %He first argued with ``Young Hegelians'' that their views were ``ideological'' and overestimated the value and role of ideas in history and social life \cite{Marx1845}. 
As noted by Thompson, Marx derived the polemical conception of ideology, ``a theoretical doctrine and activity which erroneously regards ideas as autonomous and efficacious and which fails to grasp the real conditions and characteristics of social-historical life'' (\cite{thompson2013ideology}, pp. 34-35). This ideological lens played a critical role in his socio-political analysis of production and the diffusion of ideas during the movement of capitalism. In \textit{Manifesto of the Communist Party}, Marx and Engels defined ideology as ``a system of ideas that expresses the interests of the dominant class but represents class relations in an illusory form'' (\cite{thompson2013ideology}, pp. 37). Moreover, Marx also developed the latent conceptualization of ideology as ``a system of representations which serves to sustain existing relations of class domination'' (\cite{thompson2013ideology}, pp. 41). %by orientating individuals towards the past rather than the future, or towards images and ideals which conceal class relations and detract from the collective pursuit of social changes'' . %In general, Marx's unique contribution advanced ideology theory to a higher level and influenced its future development.

Later, ideology was glowingly viewed in terms of its roots in individual and group interests, and has been adopted in many disciplines with rich interpretations \cite{thompson2013ideology}. First, some Marxists, such as Lenin and Luk\'acs extended Marx's theories, e.g., Lenin's ``Socialist ideology'', and Luk\'acs's ``Proletarian ideology'', while keeping consistent with Marx's focus on analyzing social classes \cite{thompson2013ideology}.
Ideology was also adopted by scholars from philosophy, politics, social science, etc. For example, Mannheim considered ideology as ``the systems of thoughts or ideas'' that are socially situated and collectively shared. He aimed to provide ``a revised view of the whole historical process'' with the ideological lens \cite{Mannheim1939}. In the 1950s, organization science researchers introduced ideology into their work \cite{hartley1983ideology}. Bendix first presented a notion of ``organizational ideology'' as ``all ideas which are exposed by or for those who seek authority in economic enterprises, and which seek to explain and justify that authority'' \cite{bendix2013work}. Gradually, organizational ideology had shifted to norms, beliefs, and values shared by members in an organization at both fundamental and operative levels \cite{Weiss1987}. With such a lens of organizational ideology, researchers have studied various aspects of organizations, including organizational structure, knowledge sharing, engagement, and so on \cite{gibbs2013overcoming,hinings1996values,kabanoff1991equity, maclean2014living}.

\begin{figure*}[h]
    \centering
    \includegraphics[width=.98\textwidth]{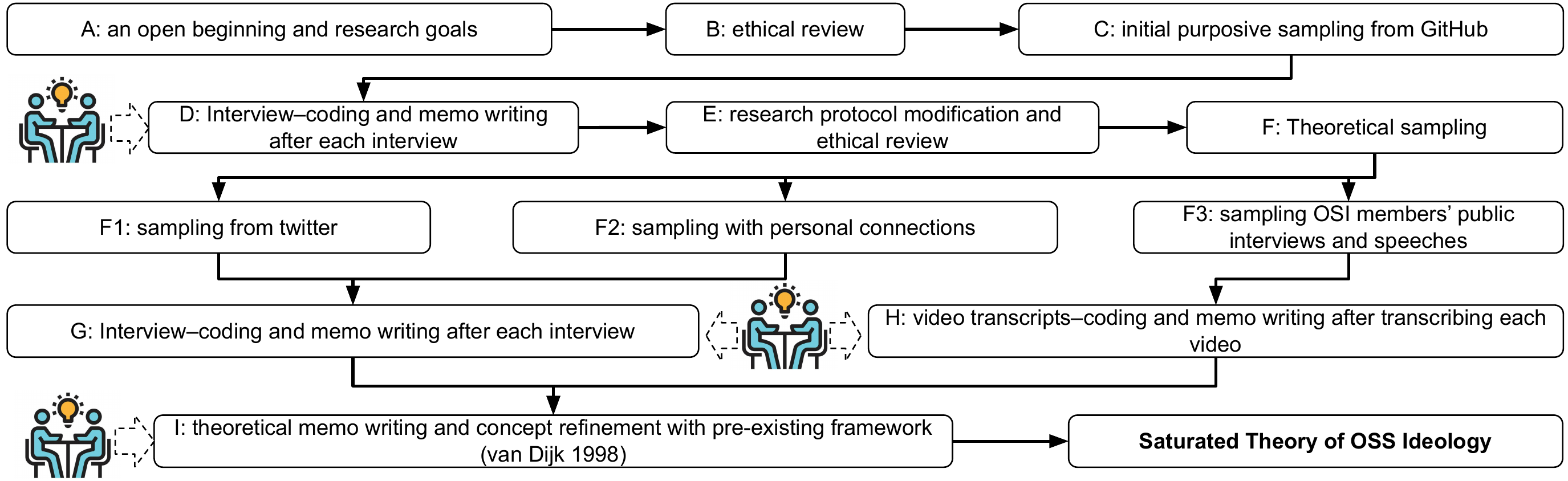}
    \caption{Overview of the research process.}
    \label{fig:method}
    \vspace{-1em}
\end{figure*}

\subsection{Exploration of OSS Ideology}

Researchers have been studying the OSS movement and exploring its ideology since it was first launched. Scholars from multiple disciplines advanced OSS ideology theories. \citeauthor{Ljungberg00} defined OSS ideology from two dimensions \cite{Ljungberg00}. One dimension was zealotry, i.e., people believe that OSS is either a new way of life, or a good way to build software. The other dimension was the degree of hostility to commercial software. %some people believe that both open source software and commercial software could co-exist, while others believe that commercial software should be replaced by open source software. %Although OSS ideology in Ljungberg's work tended to be extreme and lacked explicit definition, several norms in open source development were identified, such as gift economy, knowledge sharing, and business model. 
\citeauthor{StewartG06EffectivenessIdeology} developed the three-tenet  (beliefs, values, and norms) framework by combining famous advocators' narratives and the prior literature \cite{StewartG06EffectivenessIdeology}. For example, ``open source development methods produce better code than closed source'' is a belief in their framework. However, their framework is preliminary and far from comprehensive; even themselves admitted that.
Researchers applied these frameworks to study OSS ideology's impacts. At the project level, OSS ideology could affect the effectiveness in attracting and retaining contributors, the production of project outputs, and so on \cite{StewartG06EffectivenessIdeology}. At the individual level, the misfit of OSS ideology between contributors and projects could influence contributors' commitment \cite{DanielMCH18Misfit}. For example, the under-fit of OSS ideology--contributors embrace OSS ideology more than their projects do--would lower contributors' commitment. 
%These studies evidenced that OSS ideology could  understand the dynamics in OSS development.

Researchers also studied various aspects of OSS ideology, but in an implicit and fragile way. At the individual level, researchers first investigated the motivations to participate in open source development \cite{Bitzer06motivation, Gerosa21motivation, Krogh12motivation, Roberts06motivations}. Recently, monetary incentives' effects in motivating contributors were also examined \cite{ZhangWYZLW22Sponsorship}.   
Then, individuals' activities in OSS projects were also studied by researchers, to name a few, communication and social learning \cite{Zhou19retaining}, contributing and evaluating \cite{Ford19review, Gousios16pullrequest, Tsay14evaluating}, maintainers' invisible work \cite{GeigerHI21Maintainer}. Contributors' improper or unethical behaviors and such behaviors' impacts recently received much attention, particularly uncivil comments and toxic language \cite{FerreiraCA21Comments, QiuDetecting}.
%Pull-request model was commonly adopted in open source projects, a series of activities before and after submitting pull-requests were identified, such as checking pull-requests guidelines and checking whether similar pull-requests were processed recently \cite{Gousios16pullrequest}. When evaluating code contributions, both technical and social information was used by reviewers \cite{Tsay14evaluating}, some social signals, i.e., avatar image and newly submitted code snippet, tended to be more frequently used in the reviewing process \cite{Ford19review}. 
%uncivil comments existed when addressing arguments during code review, and these comments often negatively influenced code review by stopping further discussion \cite{FerreiraCA21Comments}. 
%Besides, individuals who work as maintainers in open source projects had more invisible work, such as supporting users, automating processes, organizing meetups, etc. \cite{GeigerHI21Maintainer} 
%Moreover, sponsorship, as a new mechanism in open source projects, could boost contributors' activities in the development process, but mostly in the short term \cite{ZhangWYZLW22Sponsorship}.
 Many project/community level phenomena had been related to ideology. As the fundamental element of ideology, values continuously received researchers' attention, for example, transparency \cite{Dabbish12transparency}, diversity \cite{Bosu19diversity, Vasilescu19diversity, Canedo20women}, and code of conduct enforcing these values \cite{LiPFD21Conduct}. Some work extended the focus to individual and group activities which are guided by values. These activities include collaboration and coordination \cite{JensenS05, KlugBH21Roadmap,MockusFH02Coordination}, licensing \cite{Colazo09license,Fitzgerald06}, governance \& decision-making \cite{Fieding99leadership,JensenS05, Moon00Linux}, and knowledge sharing \cite{Iskoujina15knowledge, Zagalsky18knowledge}. From a structural perspective, researchers deconstructed how individuals form communities. They found that different roles of contributors consist of a centralized, layer-upon-layer structure \cite{JensenS07Role}, with pathways to allow leadership to emerge \cite{Trinkenreich20roles, HergueuxK22Leadership}.

In general, the research interests related to OSS ideology were increasing \cite{Crowston12, YueYYWR21}. However, most studies tended to be knowledge fragments that focus on particular aspects, and there were few efforts to integrate them into a coherent body \cite{Crowston12}. One of the reasons could be the lack of an explicit definition of OSS ideology \cite{YueYYWR21}. Moreover, the samples in these studies often focused on high-profile OSS projects and famous technical icons, and almost half of them were skewed toward a single project \cite{Crowston12}. That potentially overlooked the voice of grassroots and non-technical contributors in the open source movement \cite{YueYYWR21}. Therefore, we first derived an explicit definition of OSS ideology and a framework to facilitate the inquiry. Then, we collected data from both first-line OSS participants and OSI board members, and built an empirical theory of OSS ideology from the collected data.

\section{Research Methods}
\label{methods}
 The study design followed the grounded theory methodology \cite{charmaz2006constructing} since it fitted our scientific inquiry well. First, there was very limited extant theory examining the ideologies in the OSS movement. Second, empirical evidence related to OSS ideologies was also lacking. Under such circumstances, grounded theory entailed the discovery of concepts from data in an inductive process \cite{charmaz2006constructing}, allowing the generation of novel and accurate insights into the topic under study. Moreover, the iterative process makes the theory falsifiable, dependent on context, and never completely final \cite{haig1995grounded}. Thus, we could approach OSS ideology by learning from OSS participants' experiences and reflections within particular contexts, rather than with the intent to measure something against pre-existing theories. Fig. \ref{fig:method} provided an overview of the whole process. 

%In general, we employed grounded theory methodology, because there is no dominant theory of OSS ideology in literature yet, and grounded theory is useful to construct such a theory \cite{muller2010grounded}. Fig. \ref{fig:method} is the overview of the research process, and we adopted an adapted version of grounded theory in our study \cite{SarkerLS01}.

\subsection{Data Sources}
 
We followed an iterative \textit{theoretical sampling} process to collect data \cite{charmaz2006constructing}. Data collection was based on the gaps in the emerging theory identified in data analysis \cite{corbin2008basics, charmaz2006constructing}. We had two data sources: (1) interviews with OSS members, and (2) video-recorded public interviews/speeches of current and former OSI board members. Both constituted legitimate data sources in grounded theory since grounded theory accommodates ``Interviews, field observations, documents, video, etc.'' \cite{corbin2008basics}. 

We did not randomly choose these two data sources. In particular, with the progress of the interviews and data analysis, we noticed that most interviewees were technical staff, and their experiences were hardly beyond the project level. Thus, such a sample was insufficient for our theory development. Meanwhile, the potential informants representing non-technical aspects of OSS (e.g., legal counsels, high-level leaders) were often hard to access directly to arrange interviews. Therefore, we resorted to secondary data sources. After careful consideration, we chose the video recordings of public interviews/speeches of OSI board members on \textsc{Youtube} and other video-sharing sites. They were often renowned contributors or highly respected individuals in the OSS world and leading proponents of the OSS movement, e.g., Josh Berkus was well known for contributions to multiple high-profile projects, e.g., Linux, PostgreSQL. This data source had three advantages. First, OSI board members consisted of a wide range of OSS participants. Many were opinion leaders, activists, and lawyers, thus forming a good complementarity with the interviews and satisfying grounded theory's requirement of theoretical saturation \cite{corbin2008basics, charmaz2006constructing}. Second, public interviews and speeches of this sample often contained rich ideology-related content and influenced many OSS participants. Third, these videos were publicly accessible and in the public domain without incurring any additional ethical risks.

\subsection{Data Collection}
\subsubsection{Participants Recruitment and Interview}
We recruited interviewees in a three-wave process. As the initial purposive sampling, the first wave (C in Fig. \ref{fig:method}) recruited participants from \textsc{GitHub} through emails. We first sampled 20 projects from multiple domains and of multiple sizes. We collected their contributors' emails ($N=540$) from their public profiles if they existed. Then we sent email invitations to these potential participants. We received 19 replies expressing willingness to join the study (response rate: 3.5\%). In the second wave, we posted recruitment ads on \textsc{Twitter} to reach broader audiences. Those who signed up were screened and selected based on their responses. However, most of them either lacked OSS experiences, or were scammers, except one eligible interviewee. The third wave focused on the research team's connections. Some research team's contacts were veteran OSS practitioners. We asked them if they would like to participate; two agreed to join. In total, we recruited 22 participants from 14 countries around the world. They had diverse backgrounds in jobs, experiences, and other demographic factors. The second and third waves, and the OSI members' video data collection ($F_1$, $F_2$, $F_3$ in Fig. \ref{fig:method}) served for the theoretical sampling process.

We had semi-structured, online interviews with the participants. Each interview was about 30 minutes. First, we asked questions about their demographics and SE background. The second part was used to collect their personal opinions and stories about the OSS movement. We asked a series of open-ended questions, such as \textit{``what is your understanding of open source?''}, \textit{``what makes you start the contribution in open source projects?''}, and so on. Based on their narratives, we asked some follow-up questions for details, clarifications, and provoking further reflections. Moreover, the participants also had the freedom to share anything relevant to OSS. Upon the participants' consent, all their narratives were recorded or noted; then, these narratives were transcribed as interview data for further analysis. In total, we collected 635 minutes of interviews.

\subsubsection{Collecting Secondary Video Data}
As shown in Fig. \ref{fig:method}, video data collection was performed after the first wave of interviews started. Note that we had decided to use OSI board members' public interviews/speeches on video-sharing sites (mostly \textsc{Youtube}) as the secondary data source. First, we compiled a list of 51 people, including both 10 current OSI board members and 41 emeritus members from OSI's website. Then, we searched online for their most recent public interviews or speeches related to open source. In total, 41 of them had such videos available online. These videos were transcribed for further analysis. 
% APPENDIX \ref{secondary_data} provided the list of all 41 OSI members' names and links to the collected video.

% \begin{figure}[h]
%     \centering
%     \includegraphics[width=\columnwidth]{Example of OSI Board Members.pdf}
%     \caption{Three screenshots of secondary data (from left to right: Hong Phuc Dang, Tracy Hinds, and Jim Jagielski).}
%     \label{fig:osi_members}
% \end{figure}

%\begin{figure}[h]
%    \centering
%    \includegraphics[width=\columnwidth]{Examples of secondary data.pdf}
%    \caption{Examples of analyzed public speeches' snapshots (left-to-right, top-to-bottom: Tracy Hinds, Leslie Hawthorn, Guido van Rossum, \& Brian Behlendorf).}
%    \label{fig:secondary_data_example}
%\end{figure}

\subsection{Data Analysis}
The data analysis process started immediately after some data was collected, and both interview data and secondary video data were treated equally. The theory emerged during data analysis and formed the basis of further theoretical sampling, so we need to go back and forth between data collection and data analysis. The iterative process of data analysis and data collection stops when \textit{theoretical saturation} is reached \cite{muller2010grounded}. 

Our data analysis started with \emph{Open Coding} when provisional codes were assigned to them based on the meanings identified by the authors \cite{corbin2008basics, charmaz2006constructing}. Concepts that delineated OSS ideology in the data emerged in this step. Some concepts were grouped together as higher-level categories, and they would be refined in the following steps. Two authors coded the data independently but met virtually almost every day to discuss the codes and resolve disagreements (marked with meeting icons in Fig. \ref{fig:method}). Thus, the notion of inter-rater reliability was not applicable; rather, the two authors came to an agreement with most codes. In some rare cases of lack of agreement on certain codes, the authors chose an inclusive attitude regarding each other's proposed codes, postponing the validation of the codes until the time when more data were collected and analyzed \cite{SarkerLS01}. Doing so helped expose each researcher's individual interpretive act as early as possible, thus avoiding potential misunderstandings and inconsistencies resulting from coders' different theoretical backgrounds to be real, long-lasting threats \cite{evans2013novice}. %Tab. \ref{tab:sample_codes} illustrates how data was open-coded, the first two were from the interview data, and the last two were from the video data.

%\begin{table*}[!h]
%    \caption{Several examples of open coding.}
%    \begin{tabular}{p{0.5\textwidth}p{0.15\textwidth}p{0.25\textwidth}}
%    \toprule
%   \textbf{Data Snippets}   & \textbf{Sources}  & \textbf{Sample Codes Generated} \\
%    \midrule
%        Membership means you have voting rights in some things of architectural direction or in setting who had who fills leadership positions. & \textbf{P1} & \textit{voting rights}, \textit{decision-making}, \textit{leadership}. <\emph{memos}>\\
%        \specialrule{0em}{2pt}{2pt}
%        But besides, I didn't put my whole career on open source, because, you know, I'm working full-time in a company. & \textbf{P20} & \textit{contribute in spare time}. <\emph{memos}>\\
%        \specialrule{0em}{2pt}{2pt}
%        Share not only the original, you know, what you were originally given, but also share your modifications, share everything so the ability to use it, modify it, and then share or distribute those changes or the original thing. & \textbf{Jim Jagielski} & \textit{sharing}, \textit{modifications}. <\emph{memos}> \\
%        \specialrule{0em}{2pt}{2pt}
%        It lets you start where others have already reached rather than having to re-implement yourself. Can you imagine creating OpenStack every time you had to do a cloud deployment for your company. Isn't it great that you can just start with OpenStack and build on it. & \textbf{Simon Phipps} & \textit{reuse source code}. <\emph{memos}> \\
%    \bottomrule
%    \end{tabular}
%    \label{tab:sample_codes}
%\end{table*}

In \emph{Axial Coding}, the identified concepts and categories were put together and examined again to find the connections or relations among them. Based on these connections or relations, we could cluster some concepts together as categories. Then we further identified the properties and dimensions of these categories. %Once the categories, properties, and dimensions are identified,
After that, we returned to the data again, and re-coded them with these concepts emerged from the data. \emph{Constant comparisons} were performed from the beginning, and throughout the entire process of data analysis. First, the emerging theory of OSS ideology was repeatedly compared with old and new data. The gaps were identified and provided guidance for theoretical sampling to collect more data. When there were no new concepts emerged from the new data, theoretical saturation was achieved. 

The \emph{Selective Coding} focuses on important categories and concepts, without considering the others. The selection was based on the frequency of codes' occurrences or the patterns they appeared in the data, e.g., some concepts that were frequently mentioned in multiple interviews might be more important. Meanwhile, since an OSS ideology theory was desired, categories and concepts less relevant to OSS ideology might not be important. The selective coding resulted in a set of critical categories, as well as an in-depth understanding of them.

Then we finalized theory building through \emph{Theoretical Memo Writing and Concept Refinement}. It was the pivotal step in grounded theory \cite{corbin2008basics, charmaz2006constructing}. With the space created through theoretical memo writing, we compared the data, codes, categories, and concepts in our analysis, and focused on the theoretical propositions linking them together \cite{muller2010grounded}. Then, we refined the emerging categories and concepts. When the collected data well supported all the emerging categories and concepts, and new data could not provide any further refinements, theoretical saturation was thus reached \cite{muller2010grounded}. Moreover, unlike some fundamentalists' grounded theory excluding any literature, we reused a pre-existing theoretical framework proposed by \citeauthor{VanDijk1998} \cite{VanDijk1998} to facilitate the theory development in this step, guided by \cite{charmaz2006constructing, muller2010grounded}. We discussed the emerging categories and concepts, sorted them, and mapped them into the framework. We paid particular attention to ensuring that the framework would be adapted to fit these categories and concepts well. Finally, the empirical theory of OSS ideology emerged from the analysis.

\begin{figure*}[!h]
    \centering
    \includegraphics[width=.98\textwidth]{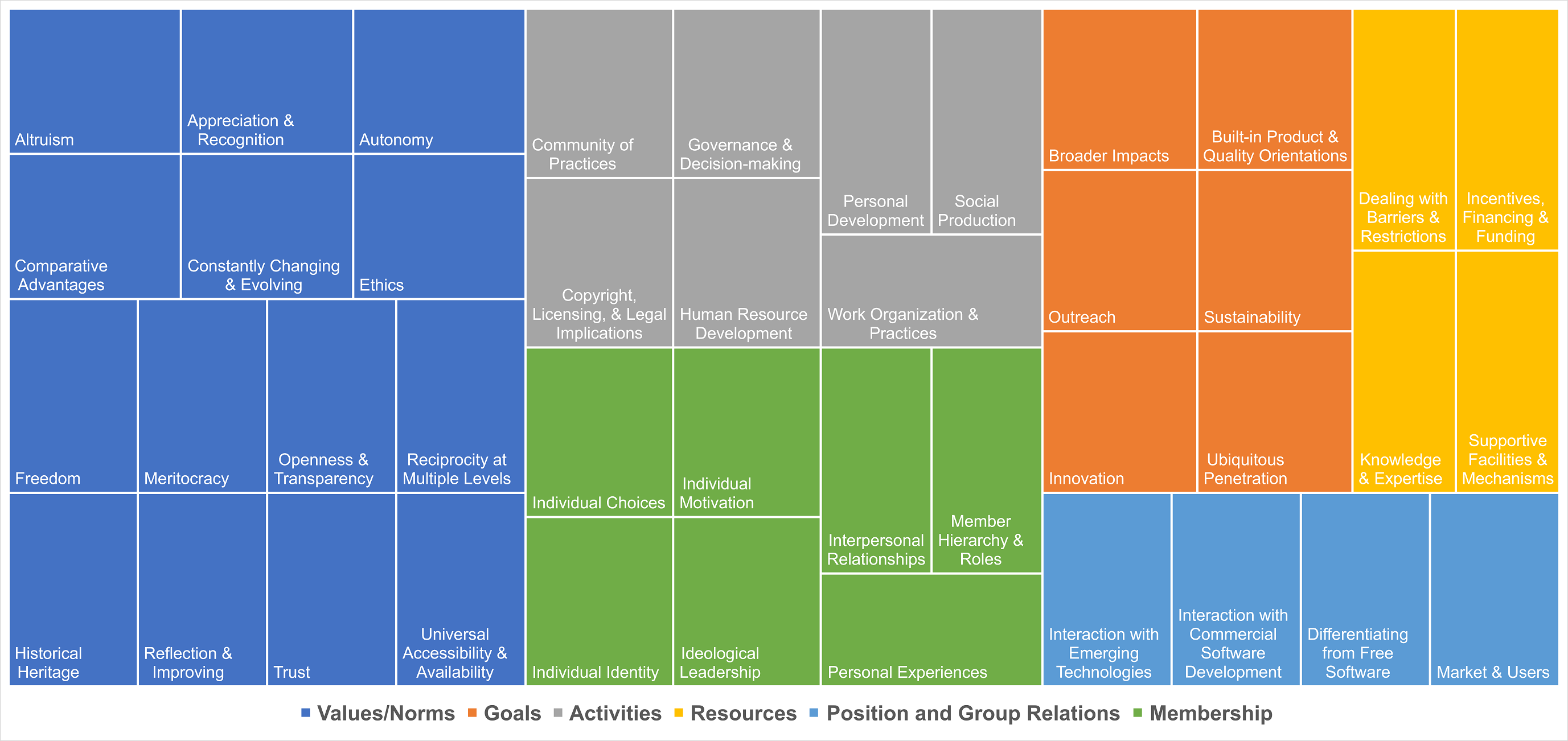}
    \caption{The overview of the empirical theory of OSS ideology.}
    \label{fig:theory}
    \vspace{-1em}
\end{figure*}

\section{The Empirical Theory}
\label{theory}
% \subsection{Overview of The Empirical Theory}
Fig. \ref{fig:theory} provided an overview of the empirical theory that emerged from the data. The empirical theory fit with van Dijk's six-category framework \cite{VanDijk1998} for ideologies, and each category contained a set of themes identified in the data analysis.

In total, the theory consisted of 42 themes in six categories. Some themes contained a series of subthemes. Due to page limitations, we could not offer detailed discussions with each of them. Moreover, it might not be necessary to do so since many of them are very straightforward and had been well covered by the extant literature, e.g., \emph{Individual Motivation} has been well-examined in \cite{Bitzer06motivation, Gerosa21motivation, Roberts06motivations, sharp2009models}. Therefore, when discussing each category, we focused on several selected key themes and subthemes that either have the potential to offer new empirical and theoretical insights or opportunities for future research. Doing so enables us to keep the paper concise without scarifying the theoretical and practical implications. 

\subsection{Membership}

%\begin{figure*}[!h]
%    \centering
%    \includegraphics[width=0.8\textwidth]{graphics/membership.pdf}
%    \caption{Themes and subthemes in the aspect of membership.}
%    \label{fig:membership}
%\end{figure*}

Membership defined the people involved in OSS, including where they came from, why and how they joined, etc. In general, ``\textit{people with the same interests}'' (\textbf{P18}) were welcome to join OSS community of their free will. Seven themes were identified in our data analysis: (1) \emph{Individual Choices}, (2) \emph{Individual Identity}, (3) \emph{Individual Motivation}, (4) \emph{Ideological Leadership}, (5) \emph{Interpersonal Relationships}, (6)\emph{Member Hierarchy \& Roles}, and (7) \emph{Personal Experiences}.  

\emph{Individual Choices} indicated that people's memberships and actions were of their own choice, which guaranteed the realization of the \emph{Autonomy} (\textbf{see \S4.2}). \emph{Individual Identity} referred to people's representation of self in constructing themselves as OSS members. \emph{Individual Motivation} reflected what motivated them to contribute, including intrinsic and extrinsic ones \cite{Bitzer06motivation, Krogh12motivation}. \emph{Ideological Leadership} was about the connection between a shared set of OSS ideologies and leadership in communities and the movement, which had been well characterized in \cite{mumford2006pathways}. \emph{Interpersonal Relationships} describe the relationships between individuals in open source development, mostly in a friendly and professional manner, while conflicts also arise in the relationships sometimes. The sixth theme \emph{Member Hierarchy \& Roles} identified the hierarchical social structure of contributors and their privileges and responsibilities defined by their roles in the membership pyramid. The last theme--\emph{Personal Experiences}--captured personal feelings in OSS development, e.g., fun experiences or being appreciated by others.

We were going to focus on two themes, \emph{Individual Identity} and {Member Hierarchy \& Roles}, for the following reasons. First, \emph{Individual Identity} organized beliefs about what a member essentially is, thus forming the foundation for the social cognition of one's membership. Second, members' identities had profound influences on all other categories. It defined their purposes, their values, and their relations to other groups, regulated what they do, and determined what resources they need. Third, \emph{Member Hierarchy \& Roles} were not only controversial for deviating from the original vision of the OSS movement \cite{demil2006neither} but also key factors directly related to governance and decision-making in an OSS community, which are perhaps the most critical activities in shaping the destiny of a project \cite{Crowston12}. Lastly, richer unanswered research questions related to them emerged from our data analysis, compared with other themes.    

\subsubsection{Individual Identity}
OSS members' identity was built on the shared understanding of their communities. However, the shared understanding was not necessarily dominating, i.e., people constructed themselves as being a member of several groups rather than a single one \cite{VanDijk1998}, and so did our informants. Many contributors developed strategies to manage multiple identities, e.g., \textbf{P9} said, ``\textit{I do not mix with the working one and the private one, and I do use separate computers..., also emails and so on.}'' However, identity struggles might be caused by one's multiple identities \cite{ashforth2008identification}. We observed such struggles among OSS contributors, particularly those sent to OSS projects by their employers. Meanwhile, other contributors were also skeptical of these company employees. \textbf{P16} shared an example, ``\textit{he is like an employee of [company], is working in the project because [company] wants to drive these projects}'' which made them ``\textit{are just as much capital as they are labor}'' (\textbf{Karl Fogel}). Therefore, interpersonal conflicts and tensions might arise between these ``employed'' contributors and others.

\subsubsection{Member Hierarchy \& Roles}

OSS community replicated the hierarchical social structures in offline human societies \cite{PintoSG16Casual} where contributors formed \emph{Pyramid Social Structure} (subtheme). Most contributors are casual contributors with limited participation, and they form the bottom layer, while ``\textit{only a few percentages of them} (subtheme), (\textbf{P4})'' would reach the upper layers. Along the path to the top, each level had different roles and associated \emph{Role-defined Privileges \& Responsibilities}. For example, maintainers have control privileges to have ``\textit{full access to everything},'' such as ``\textit{approve the change}.''(\textbf{P19}) Moreover, maintainers also need to have more commitment and take more responsibilities to keep project ``\textit{well-maintained} (\textbf{P17}).'' While such a {Member Hierarchy \& Roles} system made OSS more organized and guaranteed some continuity, it did raise ideological concerns. First, the pyramid social structure effectively created elite and non-elite social classes, which might threaten the ethical value of equity \cite{han2022cross}. Second, maintainers were increasingly put extra burdens beyond their willingness \cite{10.1145/3377811.3380920}, which constituted violations to \emph{Autonomy}. How to balance these circumstances should be addressed in future research and practices.

\vspace{-1em}
\subsection{Values/Norms}

%\begin{figure*}[!h]
%    \centering
%    \includegraphics[width=0.8\textwidth]{graphics/values-norms.pdf}
%    \caption{Themes and subthemes in the category of norms/values.}
%   \label{fig:norms}
%\end{figure*}

The literature on values/norms is vast \cite{railton2003facts}. An example is Schwartz's theory of universal content and structure of human values which defines values as guiding principles in people's lives \cite{schwartz2012overview}. In addition to the functions at the individual level, human values/norms' socio-cultural nature makes them be shared, known, and applied by members in a large variety of OSS practices \cite{Krogh12motivation}. Our data analysis revealed 14 themes related to values/norms:

\begin{enumerate} [leftmargin=.68cm, labelsep=-0.2cm, align=left]
    \item Altruism: OSS movement was driven by \emph{Altruism} because all contributors were spending their own time and effort to make something good for others voluntarily. Baytiyeh \& Pfaffman once wrote ``\emph{Open source software: A community of altruists}'' \cite{baytiyeh2010open}, which well captured such a Utopian value. 
    \item Appreciation \& Recognition: Every contribution should be recognized. People should acknowledge altruists' contributions and give them proper credit.
    \item Autonomy: OSS members were self-governed and self-determined in making choices and decisions.
    \item Comparative Advantages: OSS had \emph{Comparative Advantages} over others in its members' mind, e.g., \textbf{P16} claimed that OSS was ``\textit{more secure by having the code fully available}.''
    \item Constantly Changing \& Evolving: OSS community should be changing and evolving all the time.
    \item Ethics: Practitioners believed that open source development provides pragmatic approaches to addressing ethical concerns, and OSS communities should embrace diversity and inclusion as their essential values.
    \item Freedom: Influenced by the free software movement, OSS also values software freedom but in a more practical view.
    \item Historical Heritage: OSS should respect the historical heritages of its earlier predecessors, such as the Linux Foundation, and the free software movement.   
    \item Meritocracy: Power should be assigned according to merit only, which was determined by ``\textit{the value of your contribution} (\textbf{Jim Jagielski}).''
    \item Openness \& Transparency: The values \emph{Openness \& Transparency} were endogenous to OSS, and should be honored in every aspect of OSS. For example, OpenStack community set the Four Open principles: Open Source, Open Design, Open Development, and Open Community to ``\emph{guide everything they do}'' (\textbf{Thierry Carrez}). 
    \item Reciprocity at Multiple Levels: Reciprocal expectations were prevalent in OSS. At the individual level, \textbf{P14} shared a belief that ``\textit{how other people help me, I want to help other people}.'' At the group level, practitioners would like to ``\emph{give back to those open source projects in return}'' (\textbf{P13}).
    \item Reflection \& Improving: OSS projects should be able to self-reflect and improve their practices continuously. 
    \item Trust: OSS members should maintain certain levels of mutual trust towards each other to ensure cooperation.
    \item Universal Accessibility \& Availability. First, the source code of OSS is publicly available, and anyone can access it without asking for permission. Then, no one needs permission to use, modify, or distribute source code.
\end{enumerate}
\vspace{-0.2em}
All these values/norms had been extensively investigated in the literature, e.g., \cite{Bosu19diversity, Canedo20women, Dabbish12transparency, shaikh2016folding, StewartG06EffectivenessIdeology, Vasilescu19diversity}. Our findings generally reconfirmed theirs. Thus, we would not discuss them further and only provided examples of the more unusual ones.

\subsection{Goals}

%\begin{figure*}[!h]
%    \centering
%    \includegraphics[width=0.8\textwidth]{graphics/goals.pdf}
%    \caption{Themes and subthemes in the category of goals.}
%    \label{fig:goals}
%\end{figure*}

Goals described what members want to achieve or realize in the OSS movement. There were six themes emerged in our data analysis, which were (1) \emph{Broader Impacts}, (2) \emph{Built-in Product \& Quality Orientations}, (3) \emph{Outreach}, (4) \emph{Sustainability}, (5) \emph{Innovation}, and (6) \emph{Ubiquitous Penetration}.

\emph{Broader Impacts} captured the goals that the OSS movement would like to achieve beyond software productions. For example, \textbf{P9} mentions an ambitious vision that the aforementioned values uphold by the OSS movement ``\emph{brings humanity forward}'' in our society along with the widespread OSS. The second theme \emph{Built-in Product \& Quality Orientations} was the fundamental goal for all the other goals. Without high-quality products, none of the other goals could be realized. This goal was built in the OSS movement and considered to be a key component in Stewart \& Gosain's OSS ideology model \cite{StewartG06EffectivenessIdeology} since it directly determined many activities in development. The third theme, \emph{Outreach}, described the OSS movement's goal of promoting itself to reach a broader spectrum of users and communities. Projects utilized various channels and approaches, such as social media, to promote OSS projects. \emph{Sustainability} referred to OSS communities' goal of achieving sustainable community dynamics and growth. This goal and its antecedents have been well documented in literature such as Chengalur-Smith et al. \cite{chengalur2010sustainability, gamalielsson2014sustainability}. The fifth theme reflected that OSS development drives \emph{Innovation}. The last theme \emph{Ubiquitous Penetration} reflected the OSS movement's goal of penetrating into every aspect of modern society. Many practitioners expect that OSS would ``\emph{be foundational to modern technology}'' and thus becomes daily operation and expectations as ``\emph{a way of life}.''

Most above goals were straightforward and well-studied. Thus, we only focused on \emph{Innovation} because OSS innovation was often neglected in SE literature. Only a few scholars had worked on it \cite{feller2007open,raasch2013rise}. Moreover, even these related studies almost always focused on innovations resulting from micro-level adoption of OSS products \cite{Fitzgerald06, hinchey2016innovation}, e.g., using OSS solutions in products. The limited scope of the extant work restricted our understanding of innovation as a goal of OSS, which repeatedly emerged in our study.

Indeed, OSS aims at providing \emph{Disruptive Environment for Innovation}, which blurs the boundaries and barriers of organizations and institutions. Levine \& Prietula's four open collaboration principles-- goods of economic value, open access to contribute and consume, interaction and exchange, and purposeful yet loosely coordinated work--were all present in OSS \cite{levine2014open}. But the disruptive environment provided by OSS is much more than a playground for open collaboration. Our informants suggest a much more aggressive belief that OSS creates ``\emph{Free Market of Both Ideas and Software}'' (\textbf{Mike Milinkovich}) where people could break technical rules, explore and experiment with rich technical connotations, exercise minimal degrees of creativity for selective innovations, with very low or no cost. All indicate opportunities in future SE research in OSS innovation. Moreover, the OSS innovation's complex and multi-faceted empirical realities make its theoretical significance lie in its interdisciplinary nature. Regarding our current knowledge, the intellectual exchanges between SE and Innovation Management were limited. SE researchers interested in OSS innovations might consider integrating rich open innovation theories developed in their future research to enable cross-fertilization between disciplines \cite{davies2018neighboring}.

\subsection{Activities}

%\begin{figure*}[!h]
%    \centering
%    \includegraphics[width=\textwidth]{graphics/activities.pdf}
%    \caption{Themes and subthemes in the category of activities.}
%    \label{fig:activities}
%\end{figure*}

The category of \emph{Activities} dealt with questions such as ``what do OSS contributors do?'' and ``what are expected activities in OSS?'' This category was the most important one among the six categories. According to van Dijk (\cite{VanDijk1998}, pp. 70-71), an ideology system could be identified by one particular category. For open source ideology, its distinctions mainly lie in the activities, particularly the copyright \& licensing activities that define it (\emph{The Open Source Definition}, ver. 1.9, available at \url{https://opensource.org/osd}). Thus, OSS ideology was typically an (activity) ideology representing that OSS contributors loosely gather to form communities for producing software under specific OSS licenses. There were seven themes identified, including (1) \emph{Community of Practices}, (2) \emph{Copyright, Licensing, \& Legal Implications}, (3) \emph{Governance \& Decision-making}, (4) \emph{Human Resource Development}, (5) \emph{Personal Development}, (6) \emph{Social Production}, and (7) \emph{Work Organization \& Practices}.

\emph{Community of Practices} referred to OSS community members' collective learning activities as a group of people who ``share a concern or a passion for something they do and learn how to do it better as they interact regularly''\cite{Krishnaveni12}. \emph{Copyright, Licensing, \& Legal Implications} referred to the legal activities and practices in OSS development, particularly about dealing with copyright and patent, using OSS license, and so on. These legal activities provide explicit guarantees on the norms/values we mentioned before. \emph{Governance \& Decision-making} referred to several governance structures, i.e., benevolent dictator, walled garden, and true meritocracy, as well as the different decision-making processes that correspond to them. The fourth theme \emph{Human Resource Development} is about developing a workforce for the development, including attracting, recruiting, and retaining contributors, and facilitating their growth. \emph{Personal Development} is the activities related to an individual's professional development, such as enriching their portfolio, earning professional skills, and seeking career opportunities. \emph{Social Production} was about the collaborative nature of its members' activities which features the collective efforts of multiple entities. As its name indicated, \emph{Work Organization \& Practices} were the practices about how work is organized in OSS development, e.g., communication and coordination activities and routines of development activities.

Many above activities had been investigated, e.g. \cite{KlugBH21Roadmap, WangFWJR20Elite, CrowstonWLEH05Coordination, Zhou19retaining}. Our empirical theory was basically consistent with these literature, and we were not going to enumerate them in this section. Instead, we focused on \emph{Copyright, Licensing, \& Legal Implications} and \emph{Work Organization \& Practices}, as our informants shared additional insights leading to potential theoretical and practical implications. 

%In \emph{Copyright, Licensing, \& Legal Implications}, the first subtheme is \emph{Rationales of Dealing with Legal Issues}. Considering the majority of practitioners in open source development are still developers who have limited or zero legal backgrounds, it is impossible to require all practitioners to act like professional legal counsels. Therefore, open source license ``\emph{makes everything easier},'' and it ``\emph{talks to regular people like you and I}(\textbf{VM Brasseur}).'' Furthermore, open source license is also ``\emph{intended to be reused by people who are not lawyers} (\textbf{VM Brasseur}),'' which actually simplifies the legal issues in open source development. 

\subsubsection{Copyright, Licensing, \& Legal Implications}
There were no trade secrets in OSS since the source code must be open, no matter under which license. Thus, there should be some ways to protect intellectual property. OSS practitioners had certain \emph{Rationales of Dealing with Legal Issues} (subtheme). First, licensing was required by OSS's definition. OSI president \textbf{Simon Phipps} claimed that ``\emph{It doesn't become open source until you put an open source license on it}.'' Second, licensing is a way of explicitly offering permissions. As \textbf{VM Brasseur} argued ``\textit{Legally, if there is no license, it is automatically all rights reserved. That means the work owner gives no permission, no rights to anyone else to do anything at all}.'' Third, licensing had practical benefits by simplifying and hiding many legal details. Fourth, licenses could be the constitution of a community because contributors' participation decision implied their licensing agreements. Indeed, OSS had complicated relations with some troublesome laws, mainly on \emph{Copyright vs. License} (subtheme). Under the current copyright law, licenses offer solutions to a copyright owner to retain the rights while giving someone else the right to exercise some of them. The legal definitions of these two terms imply certain ambiguities. Meanwhile, practitioners have different ideas about who owns the copyright, what could be given to other parties, and more importantly, whether they would (partially) lose control over the work. E.g., being asked to sign a Contributor License Agreement (CLA) to transfer copyright to the project/organization often triggered some boycotts. Therefore, it is not surprising that many participants agree that there is still ``\emph{no flawless license}'' even when a large pool of legal professionals was serving the OSI board.

Licensing and other legal issues not only influenced OSS business models and operations, but also had \emph{Consequences on Development}. Licensing was powerful to enable the development and distribution models, which made OSS ``\emph{unprecedentedly spreading across all of the categories of software}'' (\textbf{Amanda Brock}). Our participants also acknowledged the significant impacts of licensing in OSS development. They mentioned something like ``\emph{license shapes community dynamic
}'', ``\emph{license governs community}'', or ``\emph{license affects project outcomes}.'' Note that such consequences are not necessarily to be positive. Former chief policy advisor of USPTO, Arti Rai once said ``\emph{law can sometimes pose, rather than resolve, problems}'' when discussing open source and legal issues \cite{rai2007open}. For example, some company-sponsored projects may use restrictive licenses to avoid competition, which de facto betrayed the OSS ideology.

Legal professionals mostly did the design of the licensing systems. Ordinary developers often had minimal legal knowledge, and as \textbf{Pamela Chestek} said, ``\emph{I don't think that most human beings are interested at all in legal instruments};'' so they might be confused about why they need to bother with legal issues. Therefore, when encountering these issues, \emph{Reluctance \& Resistance in Development} is not uncommon. Developers naturally preferred more permissive licenses. They believe that \emph{an open source license doesn't guarantee the software is good}. To solve the limitations of licensing, practitioners proposed alternative legal arrangements. For example, trademark received some attention recently (\textbf{Pam Chestek}).

\subsubsection{Work Organization \& Practices}
\emph{Work Organization \& Practices} contained three subthemes, which are \emph{Decentralized Mass Collaboration}, \emph{Communication \& Coordination}, and \emph{Regulations, Routines, and Rituals (3Rs)}. SE and CSCW researchers have spent tremendous efforts in investigating the first two subthemes since the late 1990s. However, the 3Rs received much less attention than it was deemed in SE literature, and we would focus on this subtheme.

These 3Rs were derivatives of the Values/Norms in the OSS ideology. As concrete social control mediums, they ensured the realization of Values/Norms in daily activities. First, regulations were explicit rules defining uncompromising boundaries of activities \cite{alvesson2002identity}. It helped produce appropriate individuals and excluded those who did not respect community values/norms. Regulations could be applied to both technical and social activities. For example, OSI's CFO \textbf{Tracy Hinds} explained \emph{how peer conflict regulation work to dismiss an individual do not respect community values} at the Node.js Foundation. Second, routines are a repetitive pattern of interdependent activities that can be learned and exercised by members. We observed that routines widely existed in OSS practices. Moreover, our participants also pointed out that routines might be evolving along with the development of communities and projects. In OSS, the performances of routines were continuously evaluated, which opened opportunities for changes. Thus, an interviewee (\textbf{P12}) claimed ``\emph{no routine work in open source}.'' This coincided with findings from organizational theorists such as Martha Feldman and Brian Pentland \cite{feldman2000organizational,feldman2003reconceptualizing}. Given that the dynamics of routines could be reproduced from software repositories, there were opportunities for further theory development. Third, rituals can be characterized as standardized behaviors undertaken in conditions demanding explicit expectations. 

Another vital function of the 3Rs was their channeling roles among organizational cognition of values/norms, activities, and community/project development. Studying such roles would probably lead to establishing a complete network from values/norms to products, by embedding these constructs' complex relationships. Thus, we could answer critical questions such as how value/norms were reflected in activities and project outcomes or how to leverage such channels to ensure the values/norms were honored, and hereby offer insights for making OSS more value-sensitive \cite{10.1145/3377811.3380393}.

\subsection{Resources}

%\begin{figure*}[!h]
%    \centering
%    \includegraphics[width=0.8\textwidth]{graphics/resources.pdf}
%    \caption{Themes and subthemes in the category of resources.}
%    \label{fig:resources}
%\end{figure*}

Resources were essential for a community to survive and develop. Conservation of Resources (COR) theory argues that an individual or a group shall preserve and protect those resources that they value \cite{hobfoll2018conservation}. Resources are not restricted to tangible resources but also include various intangible ones. We identified four themes: (1) \emph{Dealing with Barriers \& Restrictions}, (2) \emph{Incentives, Financing \& Funding}, (3) \emph{Knowledge \& Expertise}, and (4) \emph{Supportive Facilities \& Mechanisms}. Of these four themes, only (2) falls into the class of tangible resources; all the rest are intangible resources.

\emph{Dealing with Barriers \& Restrictions}, mentioned the resources that individuals and a community used to deal with barriers \& restrictions they faced. A typical individual-level resource is some personality traits, e.g., resilience, since they could help people deal with burnout and frustrations, particularly in one's early career phase \cite{SteinmacherGCR19}. The second theme is about money. While contributions to open source are voluntary, healthy cash flow is still quite important for many projects, especially the large ones, to maintain community infrastructures, e.g., paying for project communication services \cite{ZhouWKHU22}. The third theme was about the knowledge and expertise needed. Moreover, they were not only cognitive or epistemological but also involved in many social dimensions \cite{Mannheim1939}. According to Michel Foucault, power is based on knowledge and makes use of it; on the other hand, power reproduces knowledge by shaping it in accordance with its anonymous intentions \cite{foucault1990history}. The dynamics between \emph{Knowledge \& expertise} and power in OSS should be well worth some future investigations. The last theme refers to the supportive facilities and mechanisms implemented by open source communities, which has been well documented in literature \cite{Gousios16pullrequest, SteinmacherGCR19}. Next, we are going to focus on the second theme. It was chosen because it was often overlooked or purposefully ignored in contrast to voluntary contributions.

First, \emph{Project Donation \& Sponsorship} (subtheme) was the major way to fund an open source project, if not the only one, because most OSS projects had no revenue stream. The financial support to the project might be viewed as commercial entities' obligation to OSS communities. \textbf{Leslie Hawthorn} once said, ``\emph{if there wasn't this core infrastructure available as free and open source software... Scale up to a huge number of servers if they're paying a per license cost for every single one of them, that just would not have been possible}.'' OSS helped many companies save substantial costs. Therefore, paying back to open source was not only a philanthropic action but also a way to engage in a mutually beneficial endeavor. Nevertheless, there are no effective mechanisms to force a company to offer donations or other forms of funding. As \textbf{P1} comment ``\emph{companies that use open source often get to it because it is free... But they are not obligated to give back even when they are making a lot of money with that}.'' Therefore, many projects lacked reliable funding. Indeed, the donations are not necessarily to be large; our informants told us that even ``\emph{small donation is a great booster}.'' Besides, many participants mentioned that they would like to keep a vigilant attitude to company donations because ``\emph{money backs up decision}.'' If a project overly relies on the donation from a company, it would be inevitably influenced by the donor, which is probably not aligned with its own objectives and interests.

Providing financial support to individuals (subtheme: \emph{Financing Individual Developers}) was considered to be legitimate for certain contributors. For instance, if someone worked full-time for open source and without other means to support their basic life needs, receiving money as their living income would be acceptable and appropriate. However, \emph{Monetization of Contribution as Incentives} (subtheme), i.e., ``providing monetary incentives for contributions'' was much more controversial. There were opposite opinions on this issue. Incentive supporters argued that ``\emph{lack of incentive limits the prosperity of project} (\textbf{P17})'', and ``\emph{incentive to have higher quality work} (\textbf{P15}).'' In contrast, more practitioners held neutral to negative views of money incentives beyond supporting contributors' basic needs. Most people agreed that ``\emph{money incentive changes intention of contribution}. (\textbf{P7})'' Even those who were neutral to incentive schemes also agree that ``\emph{incentives should be within boundaries}. (\textbf{P13})'' Unfortunately, drawing such boundaries is non-trivial.

Some influential figures, e.g., \textbf{Eric Raymond}--the author of \emph{The Cathedral \& the Bazaar}, emphasized that the current supports for individuals should be reformed because ``\emph{no money goes to people maintaining critical infrastructures}.'' Those load-bearing people often worked full-time for OSS. But current OSS operations ``\emph{fail in funding the people institutionally}''. He proposed a decentralized patron system to fund these developers. In such a system, not companies but ordinary people who have regular jobs provide small amounts of money gift (e.g., \$30 per month) to load-bearing open source contributors directly. The decentralized patron mechanism, if well executed, might help solve some problems in funding open source and some individual developers in need. There have been several experimental platforms for decentralized funding, for example. But almost all of them are \emph{incentivizing/rewarding work} rather than providing living income to people in need. 

To sum up, there were many unsolved and controversial issues in establishing funding systems for OSS projects and contributors. Researchers and practitioners should pay extra attention to financial issues and develop innovative solutions to ensure the financial sustainability of projects and load-bearing individuals.

\subsection{Position and Group Relations}

\emph{Position and Group Relations} dealt with a series of questions such as, \emph{what is our social position? who are our enemies, our opponents? who are like us, and who are different?} For this category, there were four key themes: (1) \emph{Interaction with Emerging Technologies}, (2) \emph{Interaction with Commercial Software Development}, (3) \emph{Differentiating from Free Software}, and (4) \emph{Market \& Users}. The first theme reflected the fundamental positions of OSS in enabling emerging technologies (e.g., AI and Cloud) and forming Internet-wide infrastructures together with these technologies. The second and third themes were OSS' relationships with commercial software and free software. The last theme \emph{Market \& Users} summarized the relationships among OSS, market, and users from multiple perspectives such as economics, management, HCI, etc., which were consistent with findings in the extant literature \cite{bagozzi2006open, CoelhoWhy2017, Fitzgerald06, lerner2002some}. Not surprisingly, our participants and many OSI board members were most concerned with the second theme. In this section, we will focus on (2) and (3) since they are critical in answering the above position and group relation questions by relating OSS to the other two software types. Also, examining the group relations also provides instances and extensions to the prior literature on the ideological misfit between OSS communities and companies \cite{DanielMCH18Misfit}.

\subsubsection{Interaction with Commercial Software Development}
Proprietary software produced in commercial development has a lot to do with OSS. To some degree, they were enemies. However, the relationships were much more complicated, even only from the ideological lens. \emph{Coexisting, Competing, \& Replacing} (subtheme), represented practitioners' diverse views. Some believed most proprietary software already had OSS counterparts, so the commercial offerings could eventually be \emph{replaced by} OSS. A less optimistic view is that these two types of software are competing with each other. There is ``\emph{room for proprietary software} (\textbf{Brian Behlendorf})'', so it is still too early to conclude who will dominate the market in the long run. A few others hold the opinion that they could co-exist. Interestingly, those practitioners' narratives often put open source in an inferior position by downplaying OSS as ``\emph{non-core business}'' or something similar.

Meanwhile, there was a consensus that the participation of commercial companies was not motivated by ideological reasons, but practical benefits, which was consistent with Wagstrom's findings of vertical integration of commercial and OSS \cite{wagstrom2009vertical}. For a company, whether and how to involve in OSS were completely strategic decisions for business purposes. We used the second subtheme, \emph{Involving Open Sources as Strategic Decisions}, to refer to it. Our informants discussed dozens of such benefits, including branding and promotion on the market, cost-reduction by reusing and trimming down product portfolios, diversifying assets, experimenting with innovations, and hiring, among many others.

Commercial entities' involvement had profound impacts on OSS practices, and might reshape the ideologies subtly. Subtheme ``\emph{Positive \& Negative Impacts Combined}'' captured this point. Prior literature often emphasized the positive impacts of a company's engagements \cite{10.1145/3377811.3380376}, which was also confirmed by our informants. They agreed that companies ``\emph{have more resources to support}'' or ``\emph{enforce code review}''. However, many participants expressed some worries. These impacts were considered suspicious because they believed that commercial OSS's objectives differed. Moreover, companies might behave improperly in OSS communities, hence causing members' antipathy. \textbf{P16} told us that ``\emph{companies manipulate open source project} to maximize their benefits. Others mentioned companies' misbehavior, such as using a bait-and-switch tactic and treating community contributors differently.

\subsubsection{Differentiating from Free Software}
Another group relevant to the OSS movement is the free software from which the latter emerged. The relations between them were much simpler. From the historical perspective, as \textbf{Luis Villa} claimed: ``\textit{there is a lot of overlap [between free software and open source software]}.'' For instance, they both shared the idea of making source code available, and most OSS was free to use. However, they stand for views based on fundamentally different values \cite{Stallman09}. The free software movement is an ethical imperative, while OSS was a pragmatic approach to ethics (\textbf{see \S4.2}). In \textbf{P1}'s opinion, the OSS movement is about ``\emph{removing the extremist political dogma from the side of the free software project, and free software as an idea and making it more business compatible...}'' Therefore, they were not enemies, but compete in the ideological and philosophical views in people's mindsets \cite{Fitzgerald06}.

\section{Discussion}
\label{discussion}
\subsection{Theoretical Implications}

This study's main contribution to the literature was the development of the theory of OSS ideology. Our theory featured three characteristics: \emph{comprehensive}, \emph{contemporary}, and \emph{empirically-grounded}. 

First, our work realized the transition from \emph{fragile} pieces to \emph{comprehensive} framework. Such an advance offered references for future research involving ideological issues in the OSS movement. It also opened opportunities for future mappings with related studies, enabling researchers to quickly develop an impression of the corresponding research. Second, being \emph{contemporary} contained the meanings of two aspects. Firstly, when people thought of ``OSS ideology'', they might come to the early-days manifestos. However, after two decades, OSS had become a polymorphic movement involving millions of diverse participants. Modern theories should update these early manifestos and metaphors, and our theory could serve this purpose well. Secondly, being contemporary also means that our theory serves as a snapshot and reflects the current social reality of the OSS movement and as a referential point when examining the future evolution of the OSS movement. Third, ideological theories were conventionally developed deductively from several key principles rather than inductively from empirical evidence, which neglected the direct experiences and opinions of members of social movements, particularly the grassroots \cite{remmling2020sociology}. Being \emph{empirically-grounded} thus provided a partial remedy by improving the breadth of theoretical scope. Moreover, an empirical theory could also help avoid metaphysical arguments of ideologies and inform practitioners of some applicable solutions \cite{bianchin2020explaining}.

The theory also enhanced our understanding of many phenomena in OSS across micro-, meso-, and macro-levels. The discussions of the selected themes suggested many potential research opportunities. We listed a few as examples in the form of research questions.
\begin{enumerate}[leftmargin=.68cm, labelsep=-0.2cm, align=left]
    \item How to deal with identity struggles of members from companies sponsoring the project? How to reduce tensions and conflicts resulting from that? (\emph{Membership}, \textbf{see \S4.1})
    %\item 
    \item What are the impacts of licensing on OSS development? How to encourage ordinary contributors' involvement in licensing conversations? How to reduce the legal complications in OSS? (\emph{Activities}, \textbf{see \S4.4})
    \item How do routines evolve in OSS? How do the 3Rs' channeling functions work? (\emph{Activities}, \textbf{see \S4.4})
    \item How to balance companies' donations and projects' independence? How to design effective and efficient mechanisms to fund load-bearing contributors? (\emph{Resources}, \textbf{see \S4.5})
    \item How to ensure that commercial agencies honor rather than distort OSS ideologies in their business agenda? What should be done to compete contributors' minds with free software? (\emph{Position and Group Relations}, \textbf{see \S4.6}) 
\end{enumerate}
Note that the above research questions did not enumerate all potential opportunities inspired by the empirical theory. They also exhibit the breadth and depth of the knowledge body around OSS ideology, where interdisciplinary research become the unrivalled choice to tackle them.   

\subsection{Practical Implications}
\subsubsection{Anchors for Dealing with Ideological Issues} 
OSS practitioners had to deal with ideological issues. In our data, \textbf{Tracy Hinds} showed that projects' conflict management involved many ideological issues. However, people may overlook some critical ideological elements. Our theory provided anchors for them to fulfill such tasks. Besides, many communities would like to explicitly establish ideological guidelines for regulating individual and organizational practices and branding themselves, but in an overly simplified way. An example was the ``four opens'' of OpenStack. While the ``four opens'' contained some ideologies, many other ideological elements were ignored in the guidelines but caught in our theory. For example, \textbf{Eric Raymond} emphasized that infrastructure developers build and maintain the most critical backbones of the Internet but often have to worry about tomorrow's lunch because they have little means to monetize their work. As an infrastructure foundation, OpenInfra should make some resources for supporting these developers' basic living needs explicit, but not. With our theory, practitioners could check if their ideological guidelines include all necessary parts without missing any important ones.  

\subsubsection{Theory as An Evaluative Framework}
The empirical theory could also serve as an evaluative framework for OSS practitioners to evaluate if contributors and projects upheld the desirable ideologies and to what degrees. Researchers had shown that the misfit of OSS ideology between contributors and projects could influence open source development \cite{DanielMCH18Misfit}. Thus, such evaluations were of high necessity. Our empirical theory set a solid foundation for developing a measurement system for ideologies, whether qualitative or quantitative \cite{elkin2012terminology}. Such a measurement system could help identify the misfit, avoid potential conflicts in OSS development, and help match projects and contributors to improve productivity. 

\subsection{Limitations}

Our study subjected to some limitations. First, the concept of ideology is abstract and vague. We chose not to mention the term ``ideology'' during the interview, to avoid misleading the participants. However, they might not be able to explicitly express their perceptions of OSS ideology, and the collected data might not reflect their understanding of OSS ideology. Then, we intended to compile a diverse sample, including both grassroots OSS contributors and OSI board members, to reflect the population of OSS community. But it was still impossible to claim that this sample was exhaustive enough to represent the whole OSS community. And new concepts and categories might emerge when more sampled participants are in the study. Third, the empirical theory developed in our study only served as a contemporary understanding of OSS ideology, and it became ``outdated'' immediately after its establishment, like any other empirical theories related to social phenomena. With the evolution of OSS movement, further inquiries on OSS ideology are needed to keep the empirical understanding up-to-date.

% From the perspective of \textit{internal validity}, the authors first spent a fair amount of time together discussing and determining the interview questions, and ensured that all these initial questions are open-ended questions to elicit participants' narratives on open source development. We also chose not to mention the term ``ideology'' during the interview, to avoid misleading the participants to associate with political views. By the end of the interview, we also asked if the participants have anything to share but have not talked about yet. Those could help us collect their narratives in an unbiased way. From the perspective of \textit{external validity}, our sample included both leading proponents and grassroots contributors around the world. Although they represent a wide range of OSS practitioners, it is still hard to claim they could represent the whole open source community. Thus, our empirical theory of OSS ideology may not be directly generalizable to some open source practitioners and projects. From the perspective of \textit{conclusion validity}, we strictly follow the instructions of grounded theory \cite{charmaz2006constructing, muller2010grounded} in our study. When employing the existing framework in the analysis, we were aware that the framework should fit our data, and carefully ensured that the empirical theory exclusively emerged from our data.

\section{Conclusions}
\label{conclusion}

OSS development was not only a software development method, but also a social movement driven by OSS ideology. It was necessary to develop an understanding of OSS ideology to understand the OSS movement comprehensively. This paper reported on a study to develop an empirical theory for OSS ideology. Employing grounded theory, we developed an empirical theory of OSS ideology based on interviews with 22 OSS participants and 41 public interviews/speeches of OSI board members. Our empirical theory consisted of 42 themes organized into six major categories, i.e., membership, norms/values, activities, goals, resources, and positions/group relations. The empirical theory reflects a deep understanding of OSS across the community. Based on our empirical theory of OSS ideology, we further discussed some insights and potential research opportunities. To the best of our knowledge, this study is the first research inquiry that provided a comprehensive understanding of OSS ideology, and some reflections on the OSS movement latterly. Our study provided an ideological lens to investigate OSS development. It not only offered multi-level perspectives to examine OSS practices but also enabled knowledge discovery across multiple disciplines. Our empirical study shed light on the OSS movement and would facilitate researchers to gain deep insights into the ideological aspects of OSS.

\section*{Data Availability}
The interview data, even with de-identification techniques, cannot fully guarantee the anonymity of the participants. Therefore, we can only share this data with others who agree to maintain the participants' anonymity and obtain their own IRB approval (from their institutions) to use the data, instead of posting the data in an open repository. The videos are publicly accessible on several video sites, our supplementary materials provide links to them.

\begin{acks}
We thank all the interview participants, their insights were essential to the success of this study. We also thank the support from the Center for Organizational Research at UC, Irvine.
\end{acks}


\vspace{2em}

\bibliographystyle{ACM-Reference-Format}
\bibliography{manuscript}

\end{document}